\documentclass[%
 reprint,
superscriptaddress,
 amsmath,amssymb,
 aps,
pra]{revtex4-2}

\usepackage{graphicx}
\usepackage{dcolumn}
\usepackage{bm}
\usepackage{xcolor}
\usepackage{hyperref}

\usepackage{comment}

\newcommand{\SMA}{\footnote{See [URL will be inserted by publisher] for a detailed description of the simulation parameters}\space}

\newcommand{\SMAb}{\cite{Note1}\space}

\newcommand{\SMB}{\footnote{See [URL will be inserted by publisher] for a movie illustrating the interaction of a Doppler-boosted beam with a secondary target}\space}

\begin{document}

\title{Probing strong-field QED with Doppler-boosted PetaWatt-class lasers}

\author{L.~Fedeli}
\author{A.~Sainte-Marie}
\author{N.~Zaim}
\affiliation{LIDYL, CEA-Université Paris-Saclay, CEA Saclay, 91 191 Gif-sur-Yvette, France}
\author{M.~Th\'evenet}
\affiliation{Lawrence Berkeley National Laboratory, Berkeley, CA 94720, USA}
\affiliation{Presently at Deutsches Elektronen Synchrotron (DESY), Hamburg, Hamburg 22607, Germany}
\author{J.L.~Vay}
\author{A.~Myers}
\affiliation{Lawrence Berkeley National Laboratory, Berkeley, CA 94720, USA}
\author{F.~Qu\'er\'e}
\author{H.Vincenti}
\email[]{henri.vincenti@cea.fr}
\affiliation{LIDYL, CEA-Université Paris-Saclay, CEA Saclay, 91 191 Gif-sur-Yvette, France}


\date{\today}

\begin{abstract}
We propose a scheme to explore regimes of strong-field Quantum Electrodynamics
(SF-QED) otherwise unattainable with the  currently available laser technology.
The scheme relies on relativistic plasma mirrors curved by radiation pressure to boost the intensity of PetaWatt-class laser pulses by Doppler effect and focus them to extreme field intensities. We show that very clear SF-QED signatures could be observed by placing a secondary target where the boosted beam is focused.
\end{abstract}

\maketitle
Quantum ElectroDynamics (QED) is a very successful physical theory: it is a foundation of modern physics and has passed the scrutiny of the most stringent tests~\cite{HannekePRL2008,AoyamaPRD2018,AlighanbariNature2020}. Yet, its Strong Field regime (SF-QED)~\cite{RitusAnnalsPhys1972, ReinhardtRepProgPhys1977, BerestetskiiBook1982, KatkovBook1998} remains mostly out of reach of experimental investigation, leaving decades-old theoretical predictions unconfirmed.

Probing SF-QED is a considerable challenge since it requires electromagnetic fields of the order of the QED critical field~\cite{Sauter1931, Heisenberg1936, SchwingerPR1951} $E_S \approx 1.32 \cdot 10^{18}$ V/m, also known as the ``Schwinger field''. $E_S$ exceeds the most intense fields available on Earth by several orders of magnitude. However, it can be approached in the reference frame of a particle in relativistic motion. For an electron, a positron, or a photon with momentum $p^{\mu}$ in a field with electromagnetic tensor $F^{\mu\nu}$, the nonlinear quantum parameter $\chi$ expresses the effective field strength relevant for SF-QED: 
\begin{equation}
\chi = \vert p_\mu F^{\mu\nu} \vert /{m_e E_S}
\end{equation}
where $m_e$ is the electron mass. $\chi \gtrsim 1$ marks the threshold of a regime dominated by SF-QED effects such as high-energy photon emission~\cite{ErberRevModPhys1966} (nonlinear Compton scattering) or the decay of high-energy photons propagating in an intense background field into electron-positron pairs~\cite{BreitPR1934, ErberRevModPhys1966} (nonlinear Breit-Wheeler pair production).

Almost all experimental results obtained so far in SF-QED rely on ultra-relativistic particles interacting with strong fields ($E \sim 10^{11}$ V/m) naturally present in aligned crystals~\cite{UggerhojRevModPhys2005} so that $\chi \propto \gamma E/E_S$ ($\gamma$ is the Lorentz factor). A very recent experiment~\cite{WistisenNatComm2018} attained $\chi \sim 1$ with the 180~GeV positron beam of the SPS accelerator at CERN. However, since no existing accelerator can provide higher energy leptons, extending this technique to the fully quantum regime ($\chi \gg 1$) will not be possible in the foreseeable future without higher field values.
\begin{figure}[t!]
\includegraphics[width=1.0\columnwidth]{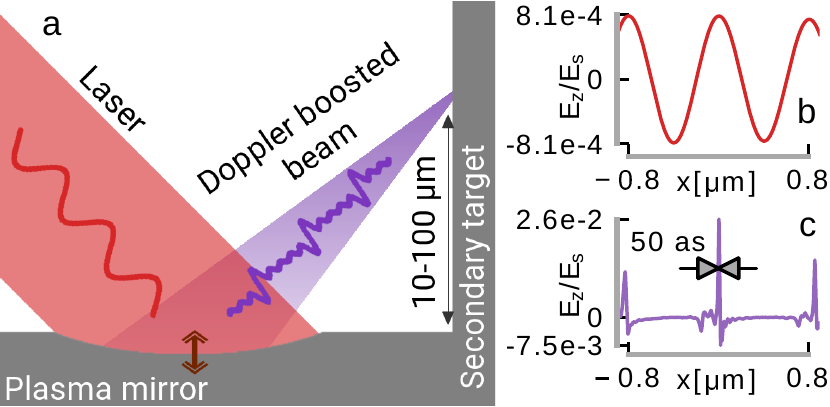}
\caption{\label{fig:sketch} Sketch of the scheme proposed in this Letter. a) A plasma mirror shaped by radiation pressure converts an intense laser pulse into Doppler-boosted harmonics and focuses them  on a secondary target, reaching extreme intensities. b,c) Electric field of a 10 PW infrared laser beam (2 $\mu$m waist) and of the generated Doppler-boosted beam at focus (both normalized to the critical field $E_S$). Note the strong change in field amplitude ($\sim \times 33$ enhancement).}
\end{figure}

\begin{figure*}[t]
\includegraphics[width=2.0\columnwidth]{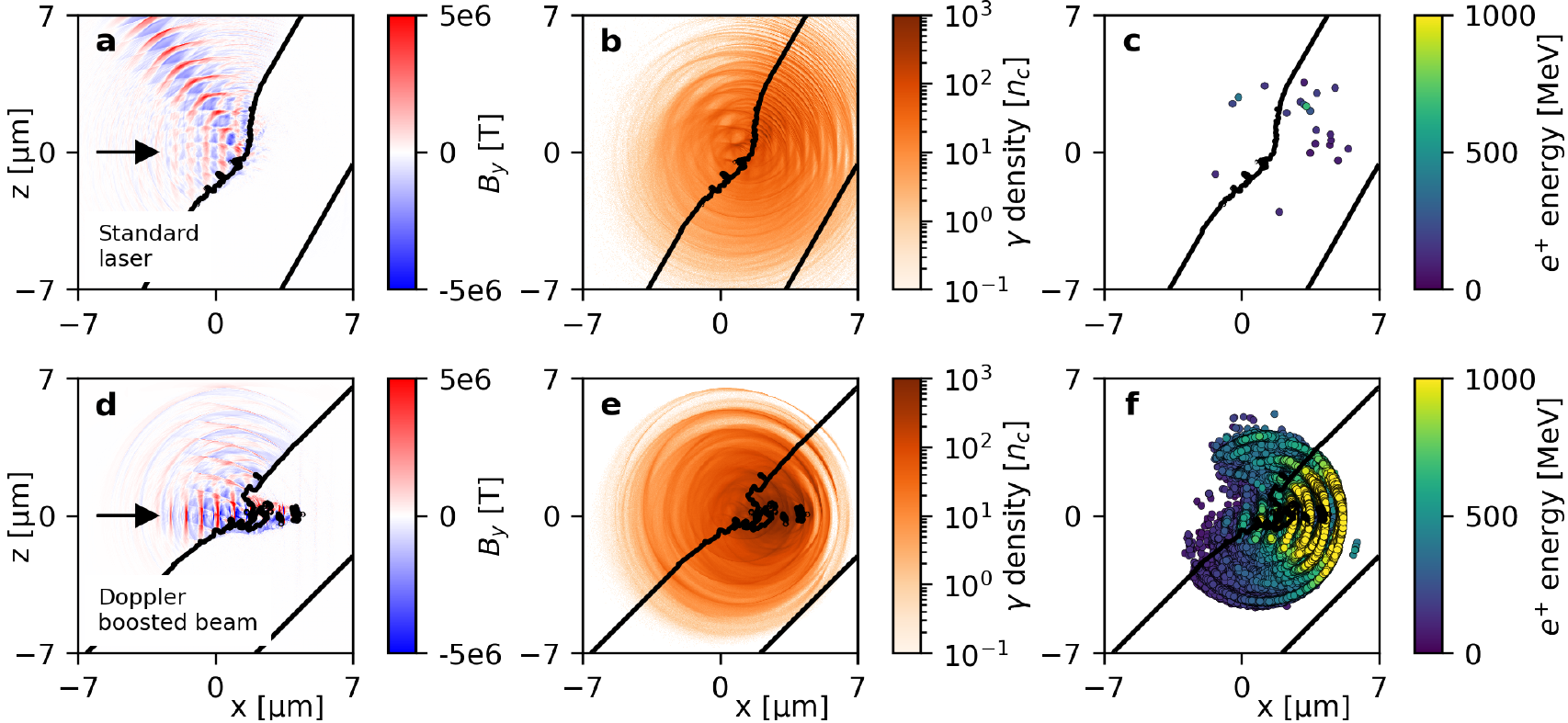}
\caption{\label{fig:sim_results} Simulation results for a 10 PW standard laser (top panels) and a Doppler-boosted beam obtained with a 10 PW laser (bottom panels), shown $\sim 10$ fs after the peak of the pulse has reached the target. a,d) Magnetic field. b,e) High-energy (E $> 2 m_e c^2$) photons density. c,f) Generated positron macro-particles. The black line marks the isodensity curve $n_e = 100~n_c$.}
\end{figure*}

Achieving stronger fields would allow the experimental scrutiny of SF-QED in a regime which has so far remained terra incognita, possibly revealing new physics beyond the standard model (e.g., Axion-like particles~\cite{KingJHEP2019}). Moreover, it could allow the generation of relativistic plasma states dominated by SF-QED effects (``QED plasmas'')~\cite{ZhangPoP2020}, which are encountered in several extreme astrophysical objects, such as blackholes~\cite{BlandfordMNRAS1977, RuffiniPhysRep2010}, pulsar magnetospheres~\cite{CurtisRevModPhys1982, HardingRepProgPhys2006, UzdenskyRepProgPhys2014, PhilippovPRL2020}, and Gamma-Ray Bursts~\cite{MeszarosAstrJ2001}. 

The highest field values available in a laboratory are currently delivered by high-power PetaWatt (PW)~\cite{Danson_HPLSE_2019} lasers, which, once focused, can deliver intensities up to $ \sim 5 \times 10^{22}$ W/cm$^2$~\cite{YoonOptExpress2019}, associated to field amplitudes of $E \sim 6 \cdot 10^{14}$ V/m. For this reason, they are emerging as a prominent path to investigate SF-QED~\cite{MarklundRevModPhys2006, BellPRL2008,ElkinaPRAB2011,RidgersPRL2012,DiPiazzaRevModPhys2012,BulanovPRA2013, WenPoP2015,MckennaRRP2016,BucksbaumLOI2020,ZhangPoP2020}. In two seminal experiments~\cite{BurkePRL1997, PoderPRX2018}, an intense laser beam was collided with a multi-GeV electron beam. Both experiments attained $\chi \sim 0.3$, reporting electron-positron pair production~\cite{BurkePRL1997} and possibly some hints of quantum corrections on radiation emission~\cite{PoderPRX2018}. To extend this scheme at higher $\chi$, large international collaborations have recently proposed two ambitious experiments: E-320~\cite{YakimenkoPRAB2019, AltarelliarXiv2019} (FACET-II, SLAC) and  LUXE~\cite{AltarelliarXiv2019,AbramowiczarXiv2019} (European X-FEL, DESY). They both aim at $\chi \gtrsim 1$, with a complex setup requiring temporal and spatial synchronization of a 10 to 100 TW-class laser pulse with the electron beam of a 10 GeV-class accelerator.

In this letter, we propose a compact scheme only requiring a single laser beam to attain a so far inaccessible regime of SF-QED. This scheme consists in considerably boosting the intensity of a PW laser pulse upon reflection off a curved relativistic Plasma Mirror (PM). We show that placing a secondary target where the PM focuses the boosted beam can lead to very high $\chi$, making the scheme appealing to study SF-QED.

A relativistic PM~\cite{ThauryNatPhys2007, VincentiNatComm2014, ChopineauPRX2019} can be formed when an ultra-intense laser beam is focused on an initially solid target (see Fig.\ref{fig:sketch}a). Upon reflection on such mirror, two processes lead to strong intensification of the reflected beam at PM focus. First, the laser field drives relativistic oscillations of the PM surface that periodically compress the reflected light energy by the Doppler effect into pulses of $\sim$ 100~as duration. These periodic temporal compressions are associated to harmonics of the incident pulse and therefore shorter wavelengths. Second, as standard high-power lasers exhibit a non-uniform spatial intensity profile at focus, the laser radiation pressure (higher at the center than at the edges of the focal spot) naturally induces a curvature of the PM surface~\cite{DromeyNatPhys2009, VincentiNatComm2014}. This curvature, along with the generation of shorter wavelengths, enables a much stronger focusing of the Doppler-boosted beam.

A recent theoretical work~\cite{VincentiPRL2019}, supported by state-of-the-art 3D simulations, proposed to leverage the combination of these temporal and spatial compression of the incident light to reach up to 3 orders of magnitude intensity gain (Fig.\ref{fig:sketch}b-c). These spatial and temporal effects induced by relativistic plasma mirrors have recently been observed experimentally~\cite{LeblancNatPhys2016, LeblancPRL2017, ChopineauarXiv2020}. If a secondary target is placed at the focus of a curved PM, the Doppler-boosted beam can accelerate its electrons to ultra-high energies. The combination of high-energy particles and strong electromagnetic fields at PM focus should result in a very high $\chi$ parameter. In practice, since such an ``optically-curved'' PM focuses the Doppler-boosted beam at distances of 10-100 $\mu$m~\cite{VincentiNatComm2014, VincentiPRL2019}, the PM and the secondary target could be the arms of an L-shaped solid target (see Fig.\ref{fig:sketch}a).

We investigated the scheme outlined above with 2D Particle-In-Cell(PIC) simulations~\cite{ArberPPCF2015}. We considered a Ti:Sapphire($\lambda = 800$~nm) laser system providing a beam with a duration of 20 fs, a peak power ranging from 1 to 15 PW, and a waist of $2 ~ \mu\textrm{m}$, which are realistic parameters for state-of-the-art laser technology~\cite{YoonOptExpress2019, Danson_HPLSE_2019}.
To quantify the benefits of our scheme for enhancing SF-QED effects we compared two configurations in which a $6~\mu$m-thick solid target is irradiated (1) directly with the focused laser beam or (2) with the Doppler-boosted beam. In configuration (2), the Doppler-boosted beam is generated by focusing the laser beam onto the first arm of an L-shaped target, and we investigate SF-QED effects occurring in the interaction of the boosted reflected beam with the second arm of this target (see Fig.\ref{fig:sketch}). 

We chose a configuration where the L-shaped target has a plasma density gradient with a characteristic length of 100 nm, which maximizes the field enhancement at PM focus~\cite{VincentiPRL2019}. The beam is focused at a distance of $\sim 15$ $\mu\textrm{m}$ down to a focal spot of $\sim 100$ nm. We also selected a laser angle of incidence of $45^{\circ}$, since it is close to optimal for harmonic generation~\cite{ThauryJPB2010}. The resulting enhancement factor is $\sim \times 33$ in field, i.e. $\sim \times 1100$ in intensity (Fig.\ref{fig:sketch}c compared to \ref{fig:sketch}b). 
 
In configuration (2), performing the complete simulation in a 2D geometry required splitting the computation in three steps: (i) Doppler-boosted beam generation on the first arm (ii) Focusing of the boosted beam with a 2D-to-3D enhancement factor to get the correct intensification at PM focus (iii) Interaction of the boosted beam with the second arm at PM focus (see supplemental materials~\SMA for details). In configuration (1), the laser pulse is directly focused on a one-armed target having a plasma density gradient with the same properties as the ones of the L-shaped target. In this case, we chose a laser angle of incidence of $30^{\circ}$, which was shown to maximize SF-QED signatures such as electron-positron pair production ~\cite{LeczLPL2020}. In both cases the target has the electron density of fully ionized plastics ($n_e = 230 ~ n_c$, where $n_c \approx 1.8 \times 10^{21}$ cm$^{-3}$ is the critical plasma density for 800 nm light).

\begin{figure}[t!]
\includegraphics[width=1.0\columnwidth]{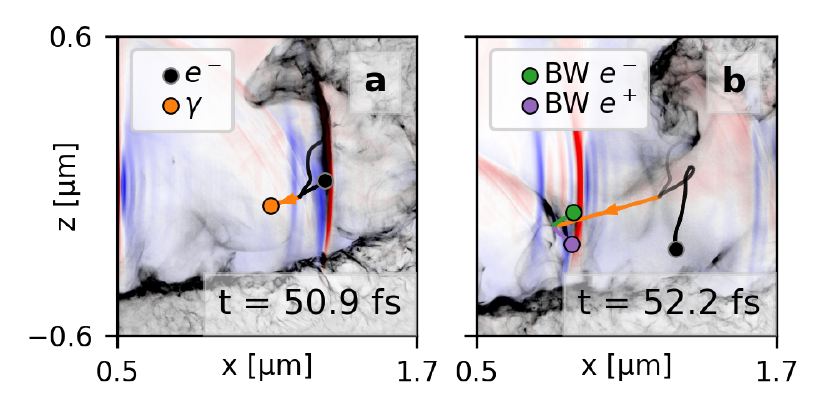}
\vspace{0cm}
\includegraphics[width=1.0\columnwidth]{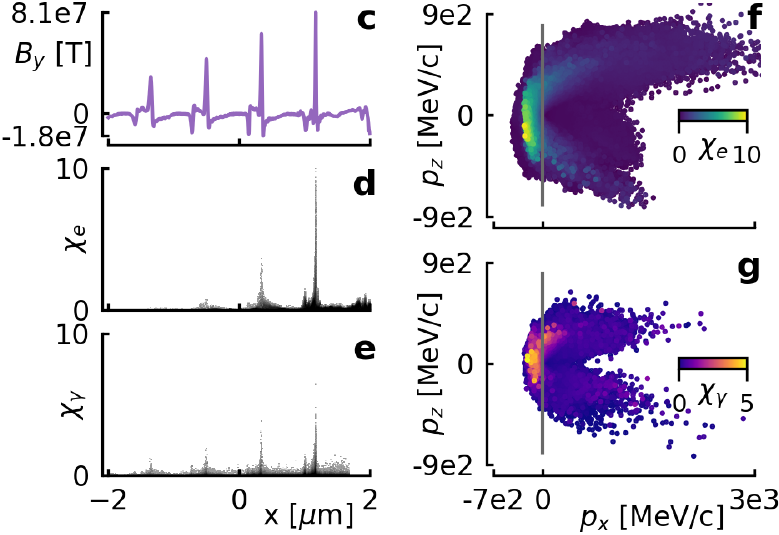}
\caption{\label{fig:mom_and_chis}
Interaction of a 10~PW  Doppler-boosted beam with the secondary target. a,b) Simulation snapshots showing respectively the emission of a high-energy photon from a target electron and the decay of this photon into a pair. Electron density is shown in gray-scale, while the transverse component of the magnetic field is shown in blue-red color scale. c) Transverse component of the magnetic field.
d,e) Respectively electron and photon phase-space projection on the $(x,\chi)$ plane.  f,g) Scatter plot of respectively electrons and photons in the $(p_x,p_z)$ plane, with $\chi$ value encoded in color. Plots in c-g) concern particles located in the interaction region, in a $4 ~ \mu\textrm{m} \times 2 ~ \mu\textrm{m}$ box, $\sim 2.5$ fs after the peak of the pulse has reached the target.}
\end{figure}
Using the WarpX+PICSAR code~\cite{VayNima2018, MyersParComp2020}, for each case we simulated a $39 ~\mu$m $\times 22~\mu$m region on the secondary target (see \SMAb for the numerical parameters). We relied on the Pseudo-Spectral Analytical Time-Domain Maxwell solver~\cite{VayJCP2013, VincentiCPC2016, BlaclardPRE2017, VincentiCPC2018} to adequately resolve the propagation of high-order harmonics. Nonlinear Compton scattering and multiphoton Breit-Wheeler (BW) pair production were taken into account with models described in~\cite{RidgersJCP2014, LobetThesis2015, LobetJoP2016} and that we optimized for GPU architectures.

The interaction of a standard laser beam with a solid density plasma at intensities high enough to observe SF-QED effects has been described in several numerical works~\cite{RidgersPRL2012, BradyPoP2014, LeczLPL2020} (see Fig.\ref{fig:sim_results}a-c). At sufficiently high intensity ($\gtrsim 10^{23} \textrm{W/cm}^{2}$) the laser accelerates target electrons to high energies, and their motion in the laser field leads to copious photon emission~\cite{ZhidkovPRL2002} via the inverse Compton process (Fig.\ref{fig:sim_results}b). Some photons have enough energy ($> 2 m_e c^2$) to decay into a pair via the nonlinear BW process while propagating in the laser field (Fig.\ref{fig:sim_results}c).

The interaction of a Doppler-boosted beam with a solid target differs substantially from the case of a standard laser, as Fig.\ref{fig:sim_results} shows. While the standard laser pulse is reflected by the solid-density plasma, the Doppler-boosted beam contains intense high-order harmonics of the original laser pulse that propagate and dig a channel in the bulk plasma (even without considering relativistic transparency~\cite{MourouRevModPhys2006, PalaniyappanNatPhys2012}, the target can only reflect harmonic orders $k < \sqrt{n_e/n_c} \approx 15$, and the Doppler-boosted beam contains intense components exceeding the $20^{th}$ order~\cite{VincentiPRL2019}). With the Doppler-boosted beam, we also observe a higher density of the emitted photons (compare Fig.\ref{fig:sim_results}e and Fig.\ref{fig:sim_results}b). However, taking into account the size of the two physical systems in the third, non-simulated, dimension (see \SMAb), the total number of generated photons is of the same order of magnitude. Finally, a striking difference is the three orders of magnitude higher amount of generated positrons (Fig.\ref{fig:sim_results}f compared to Fig.\ref{fig:sim_results}c).

Fig.\ref{fig:mom_and_chis} and the movie in \SMB allow to shed some light on the processes leading to prolific pair production with the Doppler-boosted beam. As a first step, the incident field accelerates target electrons to high energies. These accelerated electrons can emit high-energy photons in the direction of their velocity via the nonlinear Compton process (see Fig.\ref{fig:mom_and_chis}a). We observe that some of the electrons are accelerated backward (Fig.\ref{fig:mom_and_chis}f), attaining a very high $\chi$ when they cross the extremely intense attosecond pulses of the Doppler-boosted beam (Fig.\ref{fig:mom_and_chis}c and Fig.\ref{fig:mom_and_chis}d), since a counter-propagating configuration maximizes the quantum parameter. Electrons with a high $\chi$ have a higher photon emission cross-section, and they emit on average photons carrying a larger fraction of their energy. Photons emitted by these electrons also propagate backward and attain a very high quantum parameter ($\chi_\gamma$ exceeding 5) when they cross the incoming field peaks (see Fig.\ref{fig:mom_and_chis}c, \ref{fig:mom_and_chis}e and Fig.\ref{fig:mom_and_chis}g). In these conditions, since the cross-section for BW pair production grows rapidly for $\chi \gtrsim 0.5$, pair production becomes very efficient.

As the movie shows, virtually all the pairs are generated in correspondence of the field peaks, mostly ($> 90\%$) from back-propagating photons. In more than half of the cases, pair creation occurs when a high-energy photon created in an attosecond field peak crosses a subsequent attosecond peak as it propagates backward. This shows that having a train of attosecond pulses is advantageous to achieve efficient pair production. Some of the generated particles are trapped in the intense field of the Doppler-boosted beam and are accelerated forward in the plasma channel~\cite{PukhovPoP1999, JirkaNJP2020} (Fig.\ref{fig:sim_results}f), up to GeV energies.

\begin{figure}[b]
\includegraphics[width=1.0\columnwidth]{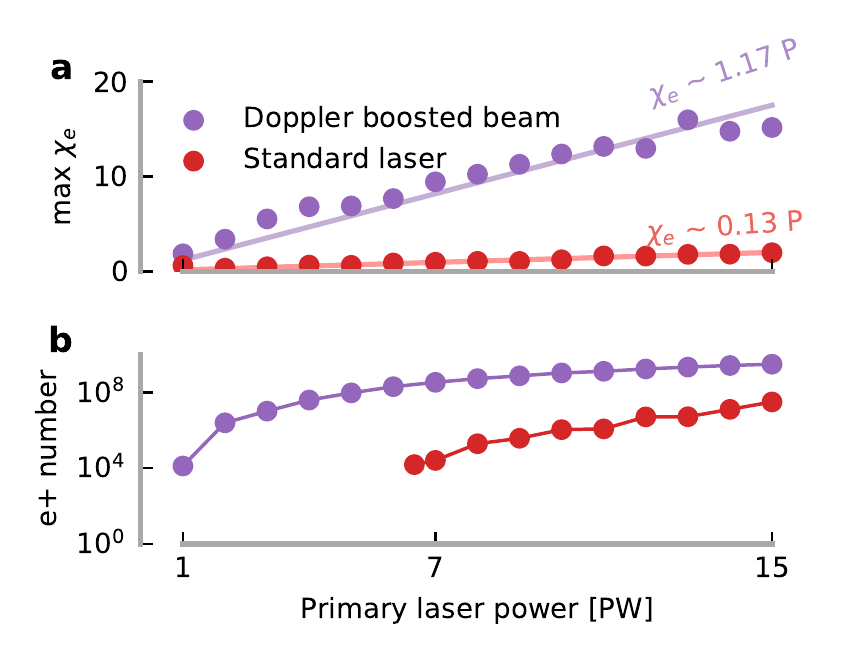}
\caption{\label{fig:max_chi} a) Maximum $\chi_e$ as a function of power. b) Number of generated positrons per laser shot as a function of laser power. For the standard laser we did not observe pair generation events below 6.5 PW.}
\end{figure}

In our simulations, we found a $\chi$ enhancement factor of approximately one order of magnitude between Doppler-boosted beams and focused standard lasers.
Indeed, Fig.~\ref{fig:max_chi}a shows that the highest $\chi_e$ reached during the interaction is proportional to the laser power and that the coefficient is $\sim 10$ times higher for the Doppler-boosted beam. We can explain this linear trend by considering that $\chi$ is a normalized product of a field amplitude and a momentum. For the standard laser, the field amplitude obviously scales with the square root of the intensity, which is proportional to laser power since we considered a constant focal spot size of $2~\mu$m. The same holds for the Doppler-boosted beam since the conversion efficiency of laser energy into harmonics depends weakly on the driving laser intensity, provided that the latter is sufficiently high~\cite{DollarPRL2013, VincentiPRL2019}. As common in laser interaction with dense plasmas~\cite{WilksIEEE1997, LiseykinaPoP2015}, we observed that electrons gain a momentum proportional to the field amplitude. The product of these terms gives an overall linear scaling with laser power.

We observed a very similar linear relation for photons (not shown here), with slightly smaller coefficients: 0.1 for the standard laser and 1.04 for the Doppler-boosted beam. Indeed, the maximum photon momentum cannot be greater than the maximum electron momentum, and they propagate in the same background field.

\begin{figure}[tb]
\includegraphics[width=1.0\columnwidth]{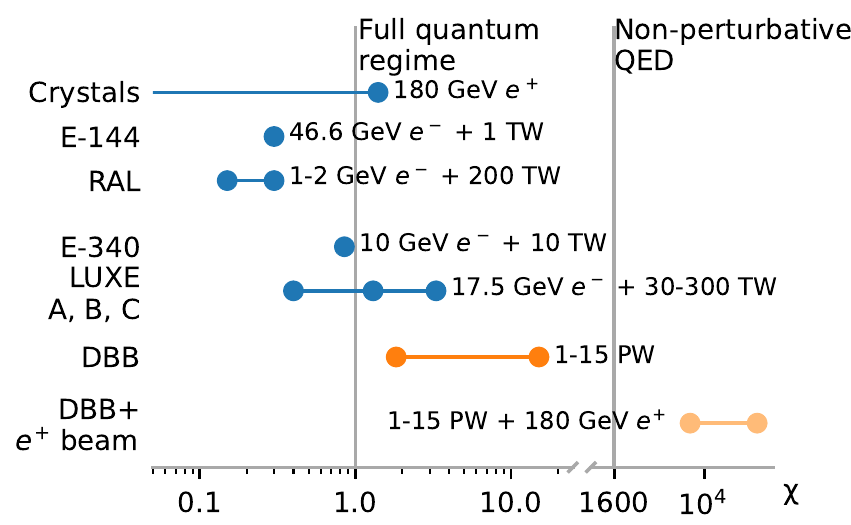}
\caption{\label{fig:allchi} Overview of the maximum $\chi$ parameter attainable with different schemes: ultra-relativistic electrons propagating in crystalline fields~\cite{UggerhojRevModPhys2005, WistisenNatComm2018}, the E-144 experiment performed at SLAC~\cite{BurkePRL1997}, the wakefield-based experiment performed at RAL~\cite{PoderPRX2018}, the E-340 experiment proposed at SLAC~\cite{YakimenkoPRAB2019, AltarelliarXiv2019}, the three phases of the LUXE~\cite{AltarelliarXiv2019,AbramowiczarXiv2019} experiment proposed at DESY, and finally the schemes based on Doppler-Boosted Beams (DBB) presented in this Letter.} 
\end{figure}
The pair production cross-section peaks at $\chi \sim 10$~\cite{ErberRevModPhys1966}. Therefore, in the parameter range that we explored, higher $\chi$ means more efficient pair production. As Fig. \ref{fig:max_chi} b) shows, for a given laser power, we obtain orders of magnitude more positrons with the Doppler-boosted beam, despite the smaller interaction volume. The enhancement is particularly significant at lower laser powers since pair generation is very inefficient for $\chi \lesssim 0.5$. These results show that, with Doppler-boosted beams, it should be possible to generate a significant number of positrons with readily available PW-class lasers. Achieving this would represent the first observation of pair production via the nonlinear BW process in a laser-plasma interaction experiment.

To conclude, Fig.\ref{fig:allchi} puts the scheme presented in this Letter in perspective with other experiments or experimental proposals devised to study SF-QED. The former would allow reaching a $\chi$ parameter exceeding that of LUXE-B and FACET-II with already operational PW-class laser systems. With soon-to-be operational 10-PW class lasers, it would even be possible to exceed the $\chi$ parameter of LUXE-C. These results mean that this scheme represents a promising complementary strategy to probe SF-QED in an unexplored regime, especially regarding QED effects in plasmas.

Finally, it's worth noting that this scheme could even be adapted to study a particularly extreme regime of SF-QED, far beyond the reach of present-day experimental capabilities. At high enough field intensities, SF-QED becomes fully-nonperturbative~\cite{RitusJETP1970, NarozhnyAPS1980, FedotovJP2017}, a regime that still  defies the formulation of a complete theory~\cite{BaumannSciRep2019, PiazzaPRL2020}. The threshold for attaining this regime is $\chi > 1600$, which is obviously a considerable experimental challenge~\cite{YakimenkoPRL2019}. However, this would be possible by coupling an optically curved PM driven by a multi-PW laser with a high-energy particle beam.
Focusing a Doppler-boosted beam on the particles in a counter-propagating configuration would result in $\chi \sim 2 \gamma E/E_S$.
Considering $E/E_S \sim 2.6\cdot 10^{-2}$ (achievable by boosting a 10~PW laser, see Fig. \ref{fig:sketch}c), the fully non-perturbative regime would be reached with a 16~GeV electron or position beam ($\gamma \sim 3.1 \cdot 10^{4}$), which is available in several accelerators worldwide. Using the 180~GeV $e^{+}$ beam provided by SPS at CERN it would be even possible to exceed the fully-nonperturbative threshold by one order of magnitude. Such experiments would be feasible with existing technology, and they would represent a promising alternative to other recently proposed even more challenging strategies~\cite{BaumannSciRep2019, BlackburnNJP2019, PiazzaPRL2020}.

\begin{acknowledgments}
This research used the open-source particle-in-cell code WarpX \url{https://github.com/ECP-WarpX/WarpX}, primarily funded by the US DOE Exascale Computing Project. We acknowledge all WarpX and PICSAR contributors.
An award of computer time (Plasm-In-Silico) was provided by the Innovative and Novel Computational Impact on Theory and  Experiment  (INCITE)  program. This research used resources of the Oak Ridge Leadership Computing Facility at the Oak Ridge National Laboratory, which is supported by the Office of Science of the U.S. Department of Energy under Contract No. DE-AC05-00OR22725. This work was supported by the  French  National  Research  Agency  (ANR)  T-ERC  program (Grant No. ANR-18-ERC2-0002). We also acknowledge the financial support of the Cross-Disciplinary Program on Numerical Simulation of CEA, the French Alternative Energies and Atomic Energy Commission. This project has received funding from the European Union’s Horizon 2020 research and innovation program under grant agreement No. 871072. The authors would like to thank Dr. Mathieu Lobet (Maison de la Simulation, CEA-Saclay, France) for useful discussions concerning the implementation of QED effects in PIC codes.
\end{acknowledgments}

\bibliography{biblio}

\end{document}